\documentclass[nature,letterpaper,twocolumn,superscriptaddress,amsmath,amsfonts,amssymb,preprintnumbers,citeautoscript]{revtex4}
\usepackage{dcolumn,graphicx,color,booktabs}
\usepackage{amsmath,bm}

\begin{document} 
\title{Anomalous High-Energy Electronic Interaction in Iron-Based Superconductor}

\author{D.\,V.\,Evtushinsky}
\affiliation{Institute for Solid State Research, IFW Dresden, P.\,O.\,Box 270116, D-01171 Dresden, Germany}

\author{A.\,N.\,Yaresko}
\affiliation{Max-Planck-Institute for Solid State Research, Heisenbergstrasse 1, D-70569 Stuttgart, Germany}

\author{V.\,B.\,Zabolotnyy}
\affiliation{Institute for Solid State Research, IFW Dresden, P.\,O.\,Box 270116, D-01171 Dresden, Germany}

\author{J.\,Maletz}
\affiliation{Institute for Solid State Research, IFW Dresden, P.\,O.\,Box 270116, D-01171 Dresden, Germany}
\author{T.\,K.\,Kim}
\affiliation{Diamond Light Source Ltd., Didcot, Oxfordshire, OX11 0DE, United Kingdom}
\author{A.\,A.\,Kordyuk}
\affiliation{Institute for Solid State Research, IFW Dresden, P.\,O.\,Box 270116, D-01171 Dresden, Germany}
\affiliation{Institute of Metal Physics of National Academy of Sciences of Ukraine, 03142 Kyiv, Ukraine}
\author{M.\,S.\,Viazovska}
\affiliation{Humboldt University of Berlin, Rudower Chaussee 25, 12489 Berlin}
\author{M.\,Roslova} \author{I.\,Morozov}
\affiliation{Institute for Solid State Research, IFW Dresden, P.\,O.\,Box 270116, D-01171 Dresden, Germany}
\affiliation{Moscow State University, 119991 Moscow, Russia}
\author{R.\,Beck}
\affiliation{Institute for Solid State Research, IFW Dresden, P.\,O.\,Box 270116, D-01171 Dresden, Germany}
\author{S.\,Wurmehl}
\affiliation{Institute for Solid State Research, IFW Dresden, P.\,O.\,Box 270116, D-01171 Dresden, Germany}
\affiliation{Institut f\"{u}r Festk\"{o}rperphysik, Technische Universit\"{a}t Dresden, D-01171 Dresden, Germany}

\author{H. Berger}
\affiliation{Institut de Physique Applique, Ecole Politechnique Federale de Lausanne, CH-1015 Lausanne, Switzerland}
\author{B.\,B\"{u}chner}
\affiliation{Institute for Solid State Research, IFW Dresden, P.\,O.\,Box 270116, D-01171 Dresden, Germany}
\affiliation{Institut f\"{u}r Festk\"{o}rperphysik, Technische Universit\"{a}t Dresden, D-01171 Dresden, Germany}
\author{S.\,V.\,Borisenko}
\affiliation{Institute for Solid State Research, IFW Dresden, P.\,O.\,Box 270116, D-01171 Dresden, Germany}

\begin{abstract}
\noindent \textbf{Strong electron interactions in solids increase effective mass, and shrink the electronic bands \cite{Grimvall}. One of the most unique and robust experimental facts about iron-based superconductors \cite{Kamihara, Stewart, Paglione} is the renormalization of the conduction band by factor of 3 near the Fermi level \cite{Popovich, Sergei_LiFeAs, Ding, Cui_NFCA, Orbitalgap}. Obviously related to superconductivity, this unusual behaviour remains unexplained. Here, by studying the momentum-resolved spectrum of the whole valence band in a representative material, we show that this phenomenon originates from electronic interaction on a much larger energy scale. We observe an abrupt depletion of the spectral weight in the middle of the Fe $3d$ band, which is accompanied by a drastic increase of the scattering rate. Remarkably, all spectral anomalies including the low-energy renormalization can be explained by coupling to excitations, strongly peaked at about 0.5\,eV. Such high-energy interaction distinguishes all unconventional superconductors from common metals.}
\end{abstract}


\maketitle

\begin{figure*}[]
\includegraphics[width=1\textwidth]{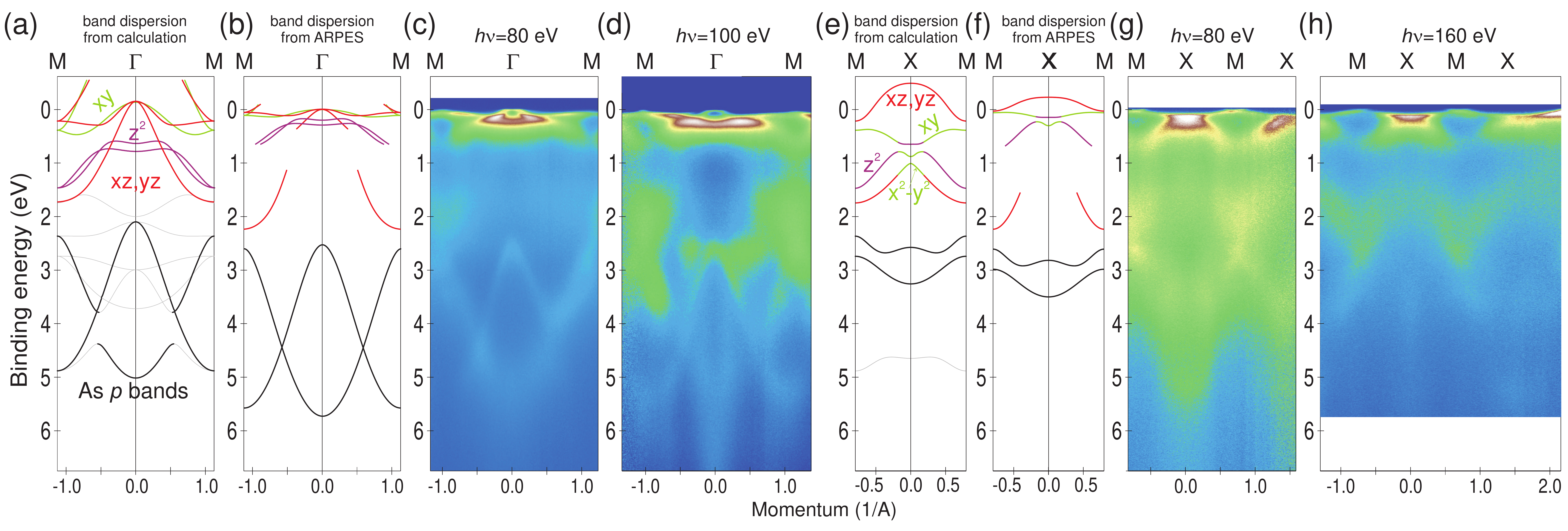}
\vspace{-0.0cm}
\caption{Theoretically calculated and derived from ARPES band dispersion in NaFeAs. (a) Calculated band dispersion in M$\Gamma$M direction. Color coding of the electronic state orbital composition is indicated in figure. Orbital polarization of the electronic states in this direction is very strong. (b) Contours of the band dispersion, extracted from ARPES intensity distribution. Extraction of experimental dispersion curves relies on the data measured at different experimental conditions, typical spectra are shown in panels (c) and (d). Panels (e)--(h) present the same information for MXM direction.}
\vspace{-0.0cm}
\label{deeeperbands}
\end{figure*}

\section{Main}
There is a fundamental problem in the condensed matter physics: Hamiltonian of any solid is extremely complicated due to large amount of involved particles\,---\,all electrons move around and act on each other. It is actually not trivial that one-electron local density approximation (LDA) gives astonishingly accurate predictions for the electronic properties of vast majority of common materials \cite{DFT, Energymin, Allen}. On the other hand, for many compounds that are currently under scrutiny of modern condensed matter physics the one-electron approach spectacularly fails, especially for those with potentially useful extraordinary properties \cite{Dagotto, Tokura}. One recent example is the class of iron-based high-temperature superconductors \cite{Kamihara, Stewart, Paglione}, where numerous experimental techniques have established that the distribution of the electronic states at the Fermi level is compressed in energy three and more times as compared to the LDA predictions \cite{Popovich, Sergei_LiFeAs, Ding, Cui_NFCA, Orbitalgap}. Since there are many candidates, not accounted for by the LDA approximation, which could lead to the renormalization at low energies and very little is known experimentally about the nature of the force capable of triple band squeezing, the problem remains unsolved. In this work we apply angle-resolved photoemission spectroscopy (ARPES) to study the electronic structure of superconducting NaFeAs covering much larger interval of binding energies than it is usually done, and show that not only electronic states at the Fermi level are renormalized, but the whole structure of the iron $3d$ band is changed with respect to LDA by a strong interaction of a particular energy scale. 



In Fig. 1 we compare the ARPES spectra recorded along the high symmetry directions in a broad energy range with corresponding results of the LDA band structure calculations for NaFeAs compound. At the bottom of the valence band, between 2 and  5\,eV binding energies ($\omega$), the photoemission intensity closely follows the calculated dispersions which correspond to the $p$-bands of arsenic. In this region there are well defined dispersion features with moderate broadening and no appreciable renormalization, implying very good agreement with theory. Near $\omega\approx2$\,eV, where the bottom of the iron $3d$-band is located, we observe the features with significantly larger scattering, but still noticeable dispersion and energy position similar to the original non-renormalized LDA bands. Comparison of the dispersions at even lower binding energies demonstrates that the experimental features change rapidly from smeared out and weakly defined to intense and well-discernible when going from 500\,meV to 0. It is clearly seen, that they are located much closer to the Fermi level than the calculations predict (compare Fig.~1(b,f) with Fig.~1(c,g) respectively). In some spectra one is able to see a kink in the dispersion at about 500\,meV [Fig.~1(c,g,h)]. A more detailed identification of the dispersive bands in the photoemission signal with the band calculation in the vicinity of the Fermi level is presented in the Supplementary Materials; basic result of such matching is that experimental bands at low binding energies are the ones predicted by LDA, but renormalized by average factor of 3.

To ensure the generality of this observation we show in Fig. 2 ARPES data taken using the light of different photon energies and polarizations. The data were recorded along various directions in the Brillouin zone. All mentioned above characteristics of the spectral weight distribution are universally present in all data sets, implying that the effect does not have a distinct momentum dependence. In addition, one can notice another common feature for the data presented in Fig. 2. It is a pronounced depletion of the spectral weight around $\omega=1$\,eV [Fig.~1(a)]. This stripe-like intensity suppression is highlighted by the dark ribbon running through all the panels of Fig. 2. It is this feature which marks the border between the region of clearly dispersing sharp band contours and the region with strong electron scattering.

Next we try to understand the origin of the above described anomalies and deviations from the one-electron picture. For a start we employ rather simple model where electrons of original LDA bands interact with a hypothetical bosonic spectrum according to the Eliashberg formalism \cite{Migdal}. Within this approach the self energy which encapsulates the many-body effects of electronic interactions is defined by the following formulas:
$\Sigma^{'}(\omega)=\int\limits^{\infty}_{0}\alpha^2F(\Omega)\ln\Bigl|\frac{\omega+\Omega}{\omega-\Omega}\Bigr| \mathrm{d}\Omega$, $\Sigma^{''}(\omega)=-\pi\int \limits^{\omega}_{0} \alpha^2F(\Omega) \mathrm{d}\Omega$, where $\Sigma(\omega)=\Sigma^{'}(\omega)+i\Sigma^{''}(\omega)$ is the self energy, and  $\alpha^2F(\Omega)$ is the Eliashberg function.

\begin{figure*}[]
\includegraphics[width=1\textwidth]{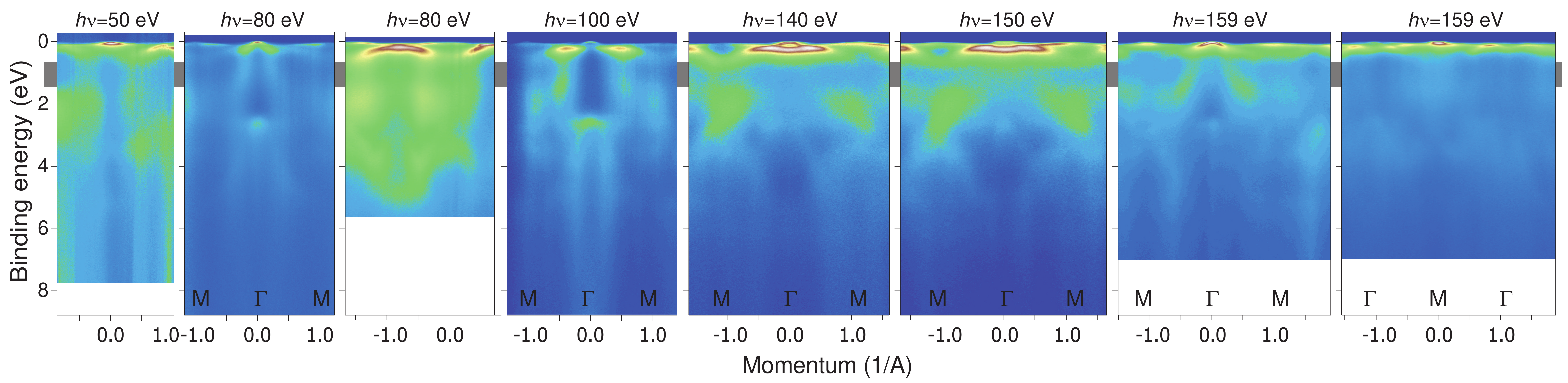}
\vspace{-0.0cm}
\caption{High-energy anomaly in spectra of NaFeAs. Spectra, recorded at different angles with various photon energies and polarizations all show well-defined dispersion at the Fermi level, large growth of scattering rate with binding energy, and a peculiar stripe of spectral weight depletion. Both spectra recorded in high-symmetry as well as off high-symmetry directions demonstrate the high-energy anomaly equally well.}
\vspace{-0.0cm}
\label{deeepermode}
\end{figure*}

In Fig.~3 we compare the calculated spectral function with experimental energy-momentum distribution of the photoemission intensity. The bosonic spectrum, $\alpha^2F(\Omega)$, was assumed to be of a simplest single-peak form. In Supplementary Materials we present a convenient analytic form for $\alpha^2F(\Omega)$, allowing for explicit integration of the expressions for $\Sigma^{'}$ and $\Sigma^{''}$. To make the comparison more transparent, the experimental conditions were chosen in such a way that photoemission matrix elements highlight one of the bands, while all others are suppressed. The model captures many important features seen in the experimental spectra. First of all, these are the sharp and strongly renormalized dispersions close to the Fermi level. The fast increase of the electronic scattering rate with binding energy at a correct energy scale is reproduced too\,---\,well-defined dispersions vanish below 0.5 eV. Finally, smeared out spectral weight is also distributed around the contours of bare dispersion. The observed broadening of arsenic $p$ bands can be accounted by the model too, provided one assumes that arsenic bands interact with the same bosonic spectrum with roughly four times weaker coupling in comparison with the one for iron bands. This result clearly implies that combining the LDA calculations with the simple treatment of the electron-boson interaction satisfactory reproduces the experimental spectral function of the whole valence band on the energy scale of up to 6 eV. Obviously, ``boson'' does not necessarily mean here a particular external bosonic excitation\,---\,it could well be just a convenient representation of purely electron-electron interactions, and it remains an open question whether corresponding bosonic excitations can be singled out.

\begin{figure*}[]
\includegraphics[width=0.63\textwidth]{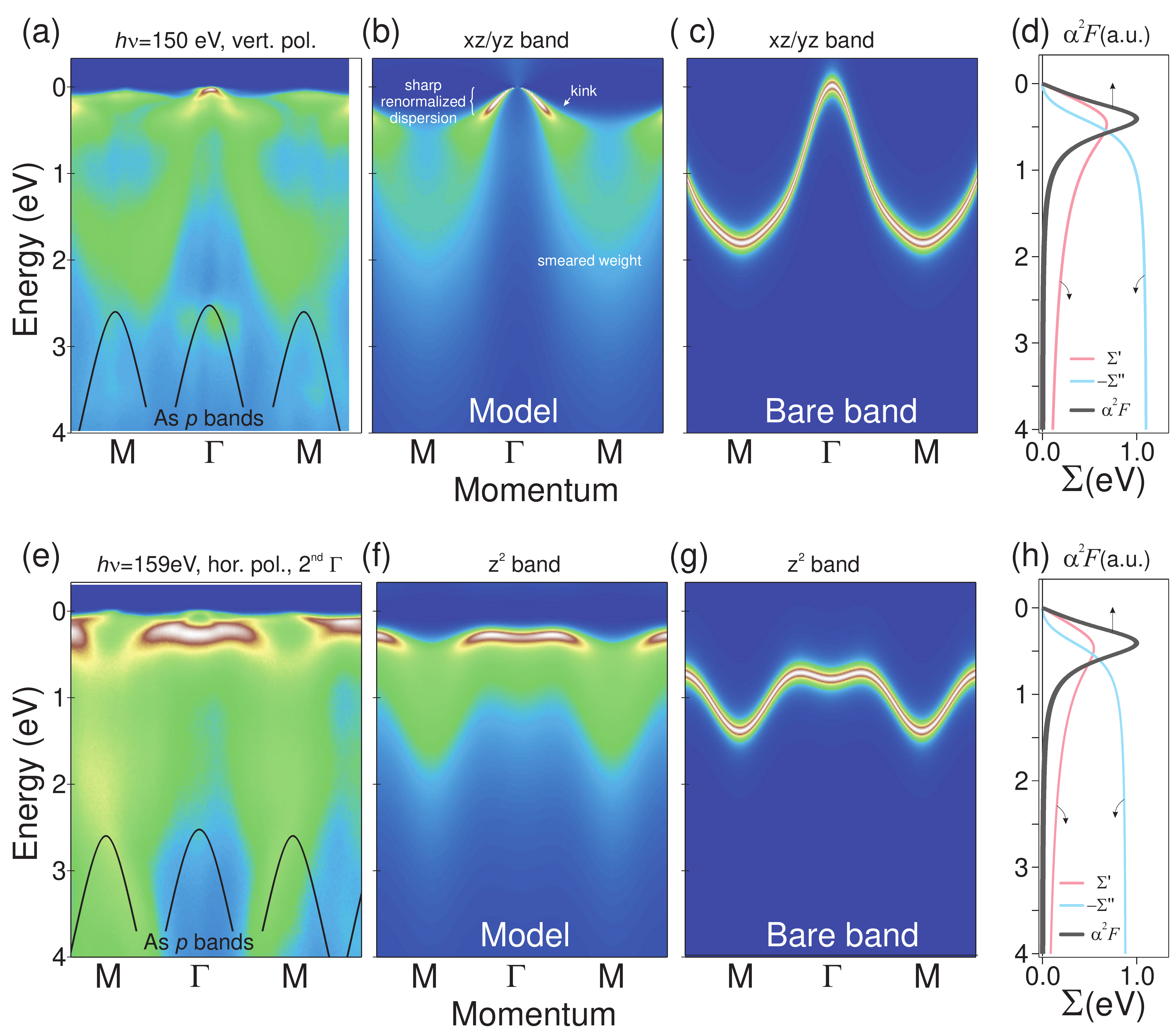}
\caption{Model for the spectral function based on the bare dispersion and electronic interaction with a hypothetical bosonic spectrum of single peak form. Experimental conditions were chosen to highlight one of the bands and suppress the intensity from all others. (a) Experimental data, recorded at photon energy of 150 eV with vertical light polarization. (b) Spectral function, obtained for lower lying $xz/yz$ band and $\lambda=2$. (c) Experimental data, recorded at photon energy of 159 eV with horizontal light polarization in the second Mahan's photoemission cone. (d) Spectral function, obtained for the $z^2$ band and $\lambda=1.6$. (e) Bosonic spectrum, $\alpha^2F(\omega)$ and self energy, used in the model.}
\label{spectral_function}
\end{figure*}

One may notice that the experimental spectral function [Fig.~3 (a), (c)] slightly deviates from the model in the region between 1.5  and 2 eV: the contours of the smeared spectral weight distribution are somewhat below the original bare bands and electronic scattering at the bottoms of these bands is somewhat lower. Both these discrepancies can be removed by allowing a \emph{decrease} of the scattering rate [blue curve in Fig. 3(e)] after it passes the maximum at binding energy of 0.5--1\,eV, which would cause a simultaneous \emph{change of sign} of the real part of the self-energy [red curve in Fig. 3(e)]. Interestingly, same effect is observed when one describes the spectra of the cuprates, ruthenates, and vanadates using the self-energy formalism \cite{Meevasana, Iwasawa, Aizaki}. Within the utilized here model such behavior of the self-energy would necessarily imply that the bosonic spectrum $\alpha^2F$ becomes \emph{negative} starting from a particular energy. This situation is directly related to the features in the spectral function, calculated for strongly correlated electronic systems \cite{Tokura, Kotliar, Held, Scala, Markiewicz}. Taken together, it means that while the intensity distribution corresponding to Fe $3d$ band, as it appears in iron pnictides and chalcogenides, can be described by weak-coupling equations surprisingly adequate, at the same time it inherits some signatures peculiar to the strong-coupling approach. Taking into account that the coupling strength $\lambda= d\Sigma^{'}(\omega)/d\omega|_{\omega=0}$ used in the model matching the experiment [Fig.~3] is around 2, we arrive at the conclusion that iron-based superconductors are essentially in the intermediate coupling regime.

\begin{figure*}[]
\includegraphics[width=0.67\textwidth]{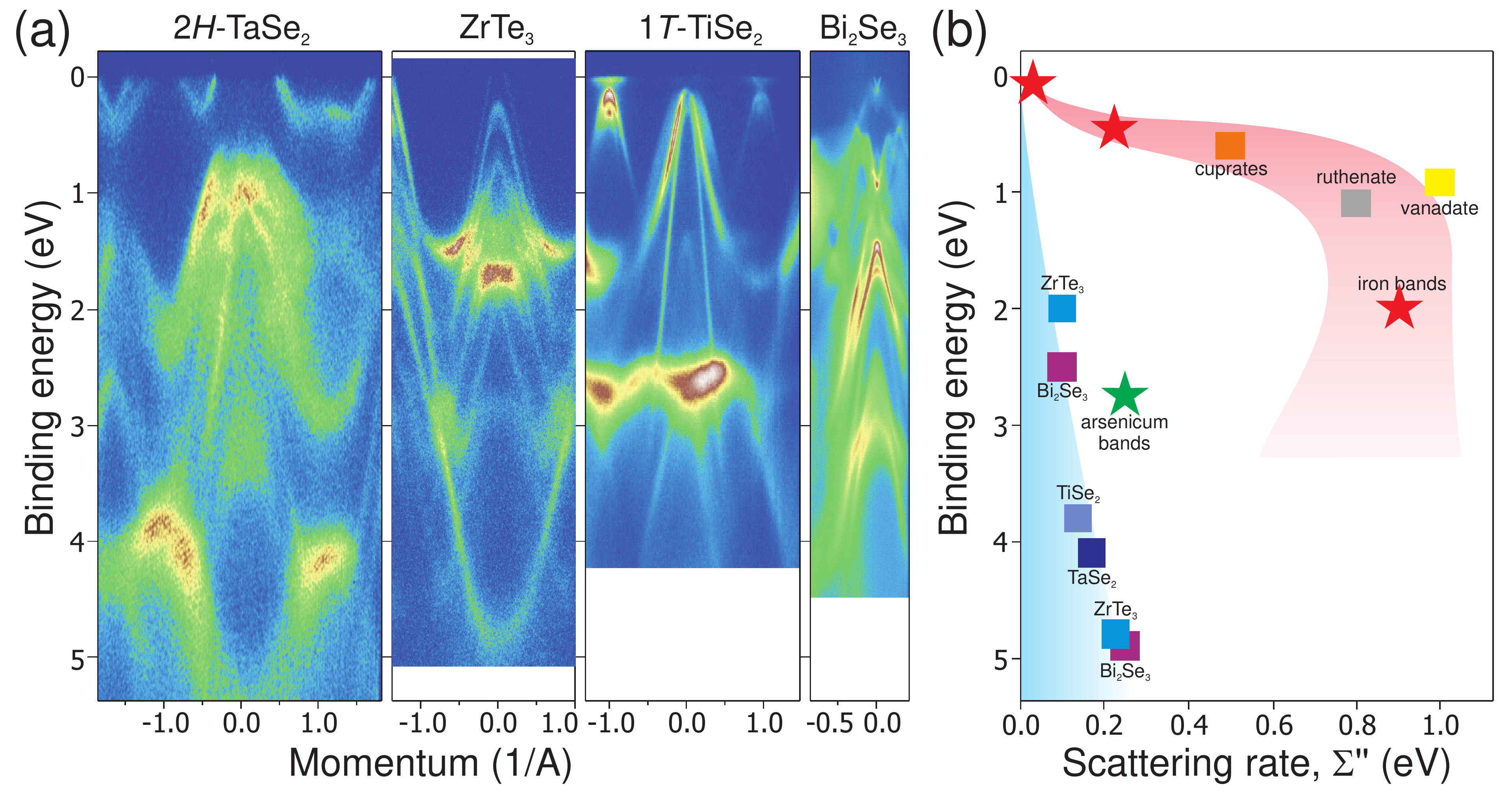}
\caption{Dependence of the scattering rate on the binding energy in different materials. In materials with relatively weak electronic interactions, like TaSe$_2$, ZrTe$_3$, TiSe$_2$, Bi$_2$Se$_3$, the spectra remain rather sharp, and the band dispersion remains well discernible down to 5 eV and further from the Fermi level. In contrast, for the materials referred to as ``correlated'', or featuring strong electron interactions, like iron-based, cuprate and ruthenate superconductors, vanadates the scattering rate growth very fast, and sharp dispersing features are observed only within several hundreds of millielectron-Volts from the Fermi level. (a) ARPES spectra of NaFeAs, TaSe$_2$, ZrTe$_3$, TiSe$_2$, and Bi$_2$Se$_3$. (b) Scattering rate, determined from the width of energy distribution curves; data for cuprates\,---\,from Ref.\,\onlinecite{Meevasana}, for ruthenates\,---\,from Ref.\,\onlinecite{Iwasawa}, for vanadates\,---\,from Ref.\,\onlinecite{Aizaki}.}
\label{All_compounds}
\end{figure*}

Observation of high-energy anomalies in the experimental spectral function such as a kink in dispersion [Fig.~1,3] and stripe of intensity depletion [Fig.~2] as well as the possibility to track intensity variations at the contours of original LDA bands allow us to determine the Eliashberg function rather precisely. What could be the physical nature of the introduced effective bosonic spectrum $\alpha^2F$? Phononic origin can be ruled out, as the typical energies of phononic modes have more than order of magnitude lower energy \cite{Boeri, Fukuda, Lee}. One of the obvious candidates would be the spin-fluctuation spectrum, as strong electron coupling to the spin resonance mode below $T_{\rm c}$, has been detected for several iron-based superconductors. However, the spin-fluctuation spectrum has maximum at about 200\,meV, and does not extend as high as 500 meV \cite{Dai_review, Fluct_pers}. Consequently, although both spin fluctuations and phonons certainly make contribution to the band renormalization at lower binding energies, they cannot be a source of the discussed here high-energy anomalies and band renormalization on the largest energy scale of $3d$ band. Among other theoretically considered possibilities the Coulomb interaction in the forms of (i) well-known on-site repulsion $U$ and (ii) recently proposed to be important in iron-based superconductors Hund's coupling $J$ \cite{Biermann, Valenti_J, Kotliar_J} are reasonable candidates for explanation of high-energy anomalies in the electronic spectrum. More theoretical work is obviously needed to understand the origin and details of the introduced here anomalous strong high-energy interaction.

It is instructive to recall that most of electronic systems are ``normal''\,---\,their spectra do not exhibit such strong anomalies and departure from LDA. We show ARPES spectra for a number of renown materials in the Fig.~\ref{All_compounds}~(a). The band dispersion can be traced down to 5\,eV binding energy and even deeper, and the agreement with LDA including the energy bandwidth is nearly perfect. A good intuitive quantity for illustration of electronic interaction strength is the scattering rate $\Sigma^{''}$. A plot of binding energy dependence $\Sigma^{''}(\omega)$ for several materials is adduced in the panel (b) of Fig. 4. There is a drastically different behaviour of the scattering rates for groups of materials with strongly- and weakly-interacting electrons.

Certainly, the most interesting question is \emph{what is the relation of the discussed above large-scale electronic interaction to the electron pairing?} It is \textit{a priori} clear that the introduced effective bosonic spectrum can hardly be considered as a pairing interaction in a conventional sense. Moreover if one makes an attempt to estimate the $T_{\rm c}$ with the parameters of the extracted spectrum,$\lambda\approx2$, $\omega_{\rm average}\approx0.5$\,eV, than with original BCS formula one arrives at $T_{\rm c}\approx\omega_{\rm average}\cdot \exp^{-1/\lambda}=3600$\,K, and at 970\,K with McMillan's expression \cite{Carbotte}\,---\,as expected one gets way too high values. On the other hand, there is a rather clear hint for importance of such electronic interactions for high-temperature superconductivity: it is present in all studied high-$T_{\rm c}$ superconductors. Finally, there is a number of materials isostructural to iron arsenides which were synthesized with complete substitution of one or more elements, e.g. BaCo$_2$As$_2$ \cite{Xu}, BaNi$_2$As$_2$ \cite{Zhou}, SrPd$_2$Ge$_2$ \cite{Timur}. In all of these materials superconductivity is either absent, or $T_{\rm c}$ is low and superconductivity is believed to be of conventional phonon origin. Remarkably, the peculiar to iron-based superconductors band renormalization at the Fermi level in all these materials is not observed. Thus, there is strong empirical indication for strong electronic interactions on the energy scale of the whole $3d$ band to be a necessary requisite for unconventional high-temperature superconductivity.





%
%
%

\section{Methods}
ARPES measurements were carried out with $1^3$-ARPES end station at BESSY II synchrotron in Berlin (Helmholtz-Zentrum f\"{u}r Materialien und Energie) on the in-situ cleaved single crystals of NaFeAs with 1.8\% of rhodium doping in iron site. The samples were cleaved at temperature around 20\,K, exhibiting shiny flat homogeneous surface. The utilized photon energies are provided in the figures. Band structure calculations were performed using the linear muffin-tin orbital method.

\section{Acknowledgments}
We thank I.\,Nekrasov, A.\,Boris, S.\,Aswartham, A.\,Charnukha and D.\,Inosov for helpful discussions, and to R.\,H\"{u}bel and M.\,Naumann for technical support. The work was supported under grants No. BO1912/2-2, BE1749/13, NAS of Ukraine (project 73-02-14) and WU595/3-1.

\section{Contributions} 
D.E., V.Z., J.M., T.K., A.K. and S.B. performed ARPES measurements, A.Y. performed LDA calculations, M.V. calculated the model spectral function, M.R, I.M., R.B., S.W. and H.B. have grown the samples, S.B. and B.B. coordinated the project.

\end{document}


\title{Supplementary materials for the article ``Anomalous High-Energy Electronic Interaction in Iron-Based Superconductor''}

\author{D.\,V.\,Evtushinsky}
\affiliation{Institute for Solid State Research, IFW Dresden, P.\,O.\,Box 270116, D-01171 Dresden, Germany}

\author{V.\,B.\,Zabolotnyy}
\affiliation{Institute for Solid State Research, IFW Dresden, P.\,O.\,Box 270116, D-01171 Dresden, Germany}
\author{A.\,N.\,Yaresko}
\affiliation{Max-Planck-Institute for Solid State Research, Heisenbergstrasse 1, D-70569 Stuttgart, Germany}
\author{J.\,Maletz}
\affiliation{Institute for Solid State Research, IFW Dresden, P.\,O.\,Box 270116, D-01171 Dresden, Germany}
\author{T.\,K.\,Kim}
\affiliation{Diamond Light Source Ltd., Didcot, Oxfordshire, OX11 0DE, United Kingdom}
\author{A.\,A.\,Kordyuk}
\affiliation{Institute for Solid State Research, IFW Dresden, P.\,O.\,Box 270116, D-01171 Dresden, Germany}
\affiliation{Institute of Metal Physics of National Academy of Sciences of Ukraine, 03142 Kyiv, Ukraine}
\author{M.\,S.\,Viazovska}
\affiliation{Max-Planck-Institute for Mathematics, Vivatsgasse 7, 53111 Bonn, Germany}
\author{M.\,Roslova} \author{I.\,Morozov}
\affiliation{Institute for Solid State Research, IFW Dresden, P.\,O.\,Box 270116, D-01171 Dresden, Germany}
\affiliation{Moscow State University, 119991 Moscow, Russia}
\author{R.\,Beck}
\affiliation{Institute for Solid State Research, IFW Dresden, P.\,O.\,Box 270116, D-01171 Dresden, Germany}
\author{S.\,Wurmehl}
\affiliation{Institute for Solid State Research, IFW Dresden, P.\,O.\,Box 270116, D-01171 Dresden, Germany}
\affiliation{Institut f\"{u}r Festk\"{o}rperphysik, Technische Universit\"{a}t Dresden, D-01171 Dresden, Germany}

\author{H. Berger}
\affiliation{Institut de Physique Applique, Ecole Politechnique Federale de Lausanne, CH-1015 Lausanne, Switzerland}
\author{B.\,B\"{u}chner}
\affiliation{Institute for Solid State Research, IFW Dresden, P.\,O.\,Box 270116, D-01171 Dresden, Germany}
\affiliation{Institut f\"{u}r Festk\"{o}rperphysik, Technische Universit\"{a}t Dresden, D-01171 Dresden, Germany}
\author{S.\,V.\,Borisenko}
\affiliation{Institute for Solid State Research, IFW Dresden, P.\,O.\,Box 270116, D-01171 Dresden, Germany}

\begin{abstract}
\noindent
\end{abstract}

\maketitle

\section{Formulas for spectral function and for self energy}

\begin{figure}[]
\includegraphics[width=1\columnwidth]{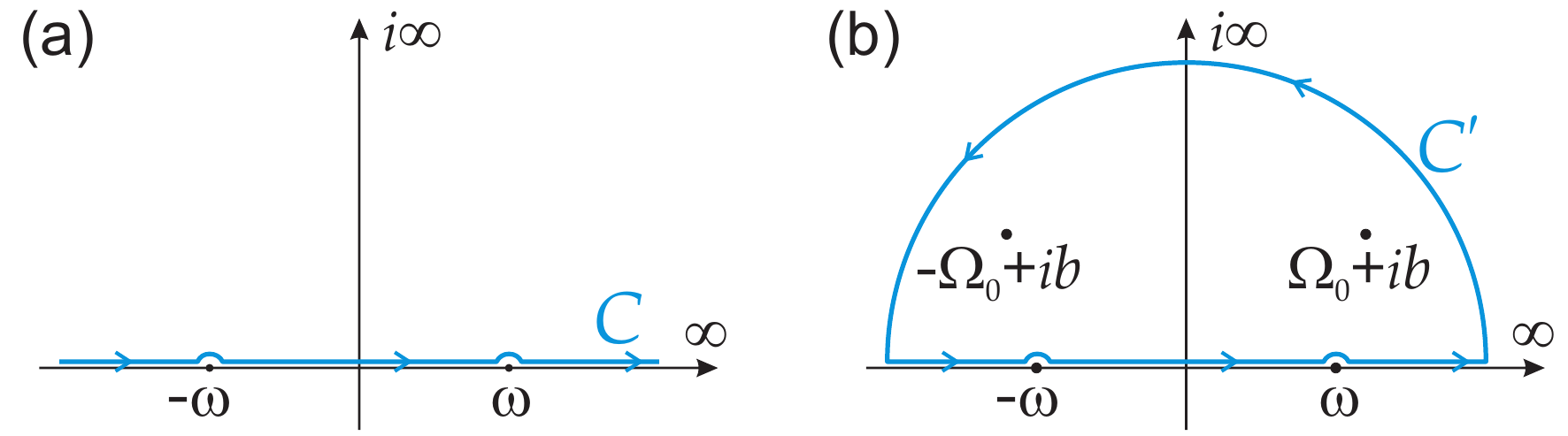}
\caption{Definition of integration paths $C$ (a) and $C^{'}$ (b).}
\label{math}
\end{figure}

\begin{figure*}[]
\includegraphics[width=1\textwidth]{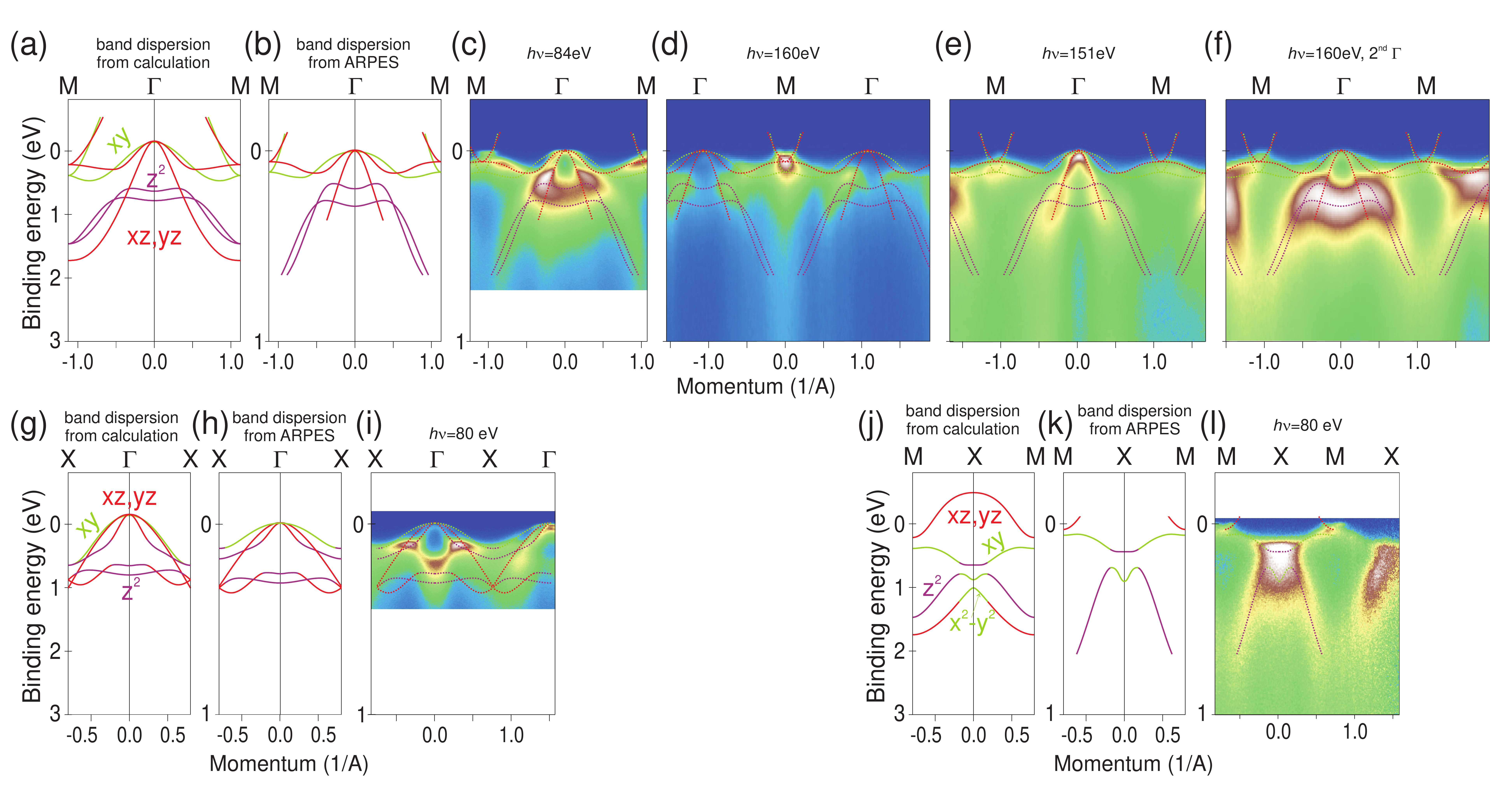}
\caption{Identification of the experimental bands in the vicinity of the Fermi level, where the dispersion is well discernible. The data are shown in the energy range of up to 1\,eV of binding energy.  Upper raw: M$\Gamma$M direction. (a) Band dispersion from calculation. (b) Band dispersion derived from ARPES. (c,d,e,f) ARPES data with superimposed contours of observed dispersion. Bottom raw: same kind of information for X$\Gamma$X direction presented in panels (g,h,i) and for MXM direction in panels (j,k,l).}
\label{ironbands}
\end{figure*}

\begin{figure*}[]
\includegraphics[width=0.95\textwidth]{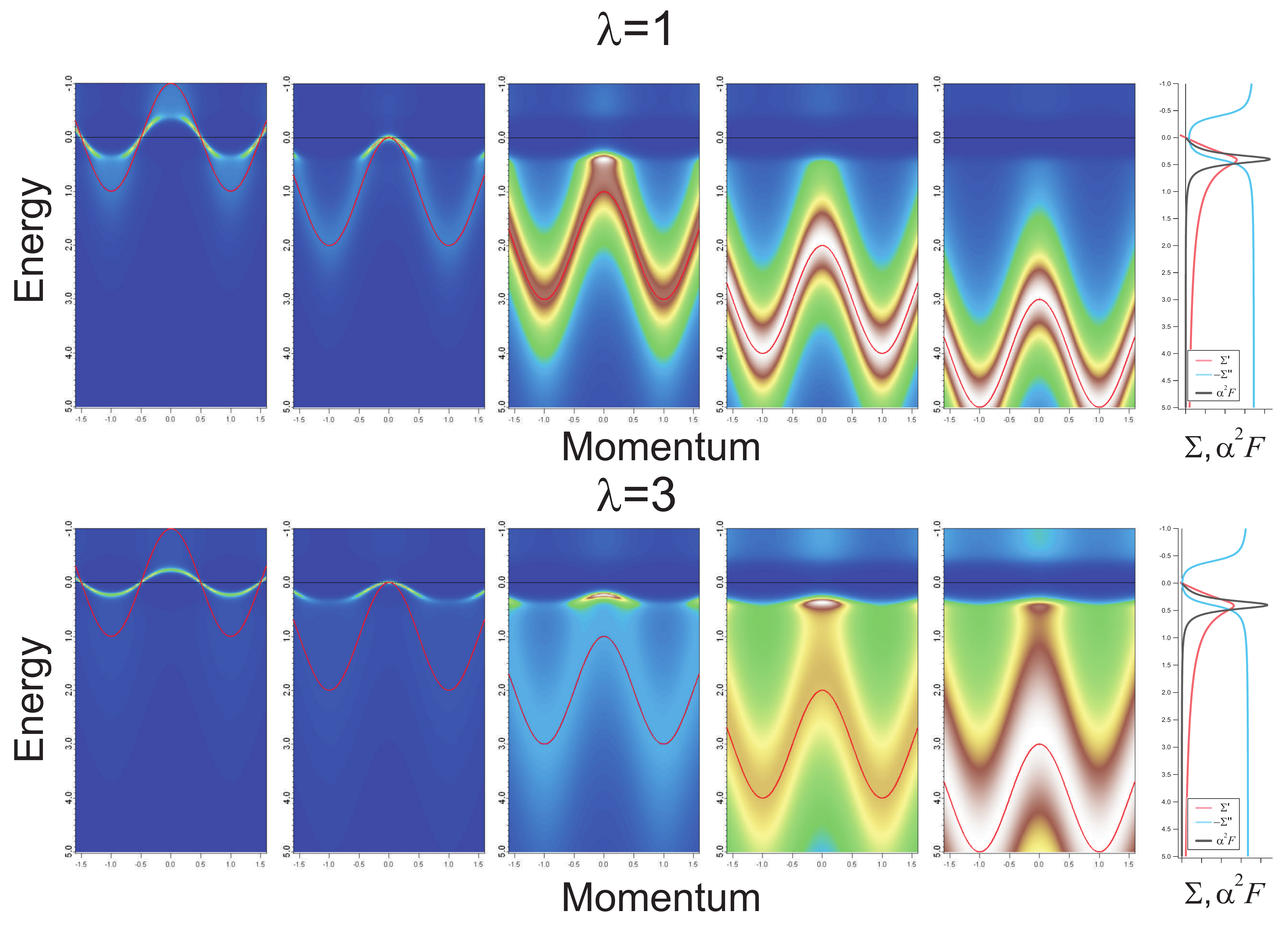}
\caption{Modeled spectral function. Bare dispersion is shown as red line. Upper raw: model for coupling strength $\lambda=1$, lower raw: for $\lambda=3$. Corresponding self energies, and bosonic spectra are shown in the rightmost column.}
\label{mutualposition}
\end{figure*}

\begin{figure*}[]
\includegraphics[width=0.85\textwidth]{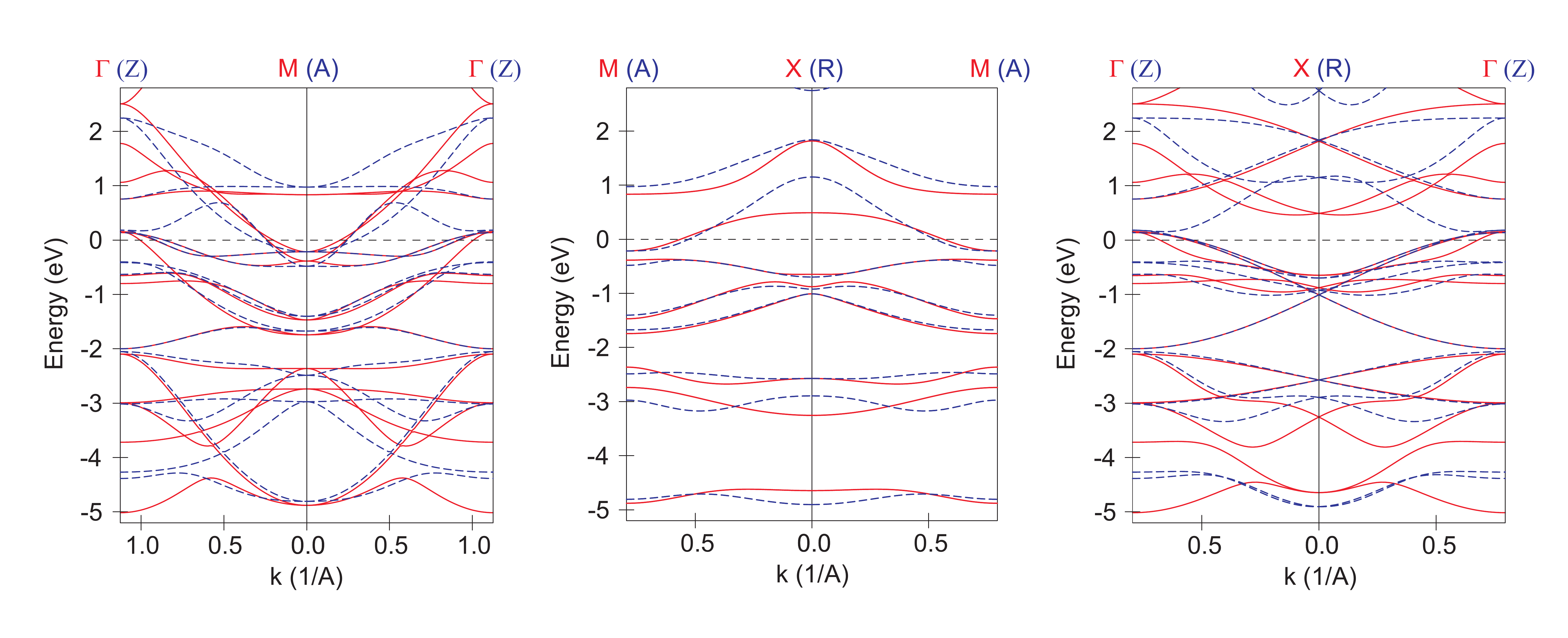}
\caption{Calculated band dispersion of NaFeAs along three principal high-symmetry directions in the Brillouin zone. Solid red lines represent the dispersion for $k_z=0$. Dashed blue lines represent the dispersion for $k_z=\pi$.}
\label{kz}
\end{figure*}

For the spectral function $A(\omega, \mathbf{k})$ we use common expression

\begin{equation}
A(\omega, \mathbf{k})= -\frac{1}{\pi}\frac{\Sigma^{''}(\omega)}{(\varepsilon(\mathbf{k}) - \omega - \Sigma^{'}(\omega))^2 + \Sigma^{''}(\omega)^2 },
\label{spectral_function}
\end{equation}
where $\varepsilon(\mathbf{k})$ is the bare band dispersion, and $\Sigma(\omega) = \Sigma^{'}(\omega) + i\Sigma^{''}(\omega)$ is the self energy. Self energy in turn can be expressed as originating from electronic interactions with hypothetical bosonic spectrum with Eliashberg function $\alpha^2F(\Omega)$

\begin{equation}
\Sigma^{'}(\omega)=\int\limits^{\infty}_{0}\alpha^2F(\Omega)\cdot\ln\Bigl|\frac{\omega+\Omega}{\omega-\Omega}\Bigr|\cdot \mathrm{d}\Omega
\label{Sigma1}
\end{equation}
and
\begin{equation}
\Sigma^{''}(\omega)=-\pi\int \limits^{\omega}_{0} \alpha^2F(\Omega)\cdot \mathrm{d}\Omega.
\label{Sigma2}
\end{equation}

Now we look for a physically reasonable shape for $\alpha^2F(\Omega)$, such that $\Sigma(\omega)$ can be expressed in a closed form. A suitable expression for $\alpha^2F(\Omega)$, representing a bosonic spectrum at positive frequencies as one peak with arbitrary height, position, and width, is
\begin{equation}
\alpha^2F(\Omega)= \frac{A}{(\Omega-\Omega_0)^2 + b^2} - \frac{A}{(\Omega+\Omega_0)^2 + b^2}.
\label{bosonic_spectrum}
\end{equation}

Integration for $\Sigma^{'}(\omega)$, Eq.~(\ref{Sigma1}), can be performed in the following way. Let us consider a function

\begin{equation}
S(\omega)=\int\limits_{C}\alpha^2F(\Omega)\cdot\ln\Bigl(\frac{\omega+\Omega}{\omega-\Omega}\Bigr)\cdot \mathrm{d}\Omega,
\label{S}
\end{equation}
where $\ln()$ denotes the principal part of the logarithm, and $C$ is given in Fig.~\ref{math}\,(a). Than we have $\Sigma^{'}(\omega) = \mathrm{Re}\bigl(S(\omega)\bigr)$. Next we consider the derivative
\begin{equation}
\frac{\partial}{\partial \omega}S(\omega) =\int\limits_{C} \alpha^2F(\Omega)\cdot\Bigl(\frac{1}{\omega+\Omega} + \frac{1}{\omega-\Omega} \Bigr)\cdot \mathrm{d}\Omega.
\label{dS}
\end{equation}
Applying Cauchy's residue theorem to the integral along the contour $C^{'}$ [see Fig.~\ref{math}\,(b)], and noting that integrand has only two poles inside $C^{'}$, $\Omega_0+ib$ and $-\Omega_0+ib$, we get

\begin{multline}
\frac{\partial}{\partial \omega}S(\omega) =2\pi i \cdot \operatorname*{Res}\limits_{\Omega=\Omega_0+ib}\biggl[\alpha^2F(\Omega)\frac{2\Omega}{\Omega^2-\omega^2}\biggr] +\\ +2\pi i \cdot\operatorname*{Res}\limits_{\Omega=-\Omega_0+ib}\biggl[\alpha^2F(\Omega)\frac{2\Omega}{\Omega^2-\omega^2}\biggr]=\\
=\frac{2\pi}{b}\biggl(\frac{\Omega_0+ib}{(\Omega_0+ib)^2 - \omega^2} + \frac{\Omega_0-ib}{(\Omega_0-ib)^2 - \omega^2}  \biggr).
\label{Cauchy}
\end{multline}
Thus, by integration the latter expression, we find

\begin{equation}
\Sigma^{'}(\omega)=\frac{\pi A}{2 b}\ln\frac{(\omega+\Omega_0)^2 + b^2}{(\omega-\Omega_0)^2 + b^2}
\label{Sigma1_formula}
\end{equation}


The integration of the equation~(\ref{Sigma2}) is much simpler, and results in
\begin{multline}
\Sigma^{''}(\omega)=-\frac{\pi A}{b}\Bigl( \arctan\frac{2\omega b}{(\omega - \Omega_0)(\omega + \Omega_0) + b^2}\\ - 2\arctan\frac{\Omega_0}{b} \Bigr).
\label{Sigma1_formula1}
\end{multline}


\section{Identification of the bands at low binding energies}

In Fig.~\ref{ironbands} we present ARPES spectra recorded in the energy window $\omega=0..1$\,eV along various high-symmetry directions in the Brillouin zone and match it with the LDA calculations. Mind that the scales for presentation of LDA and ARPES results differ three times. In spectra, recorded at different conditions, relative intensity of different bands strongly varies. For example, for the data presented in the panel (f) the intensity of the $z^2$ bands is much larger than intensity of the other bands; therefore these data are useful to track the dispersion of $z^2$ bands with no interference of other bands. Note that some discrepancy in the energy position of the $z^2$ bands in the data from panel (d), as compared to (c) and (f) can be accounted for by the moderate $k_z$ dispersion of these bands [see Fig.~\ref{kz}]. In panel (c) one can conveniently trace the upper $xz/yz$ band, as it possesses high intensity throughout the whole momentum range, and is well separated from other intense features.  Altogether the experimental data are described by the renormalized LDA bands very well: it is possible to tune the experimental conditions in such way that any chosen LDA band yields considerable intensity and vice versa all dispersive intensity has a corresponding, very close in shape, band from LDA.

\section{Dependence of the spectral function on the positions of the bare band and bosonic spectrum}

It is interesting to see how the look of the resulting spectral function depends on the mutual position of initial band and bosonic spectrum. For those bands, which lie entirely above the bosonic spectrum, e.g. for shallow $xz$ and $xz/yz$ bands in case of iron arsenides, the effects of electronic interaction with the bosonic spectrum almost entirely consist in the squeezing of the band. In contrast, for those bands, which are intersected by the bosonic spectrum, interaction effects are more interesting: one can observe sharp renormalized dispersion above the bosonic spectrum, and smeared spectral along the bare dispersion curves below the spectrum; analysis of such spectra allows for extraction of more complete information about the bosonic spectrum (see Fig.~\ref{mutualposition}).

\section{Possible broadening of ARPES spectra due to $k_z$ dispersion}

Electron energy bands in iron arsenides generally exhibit modest out-of-plane ($k_z$) dispersion. In Fig.~\ref{kz} we present contours of the LDA band dispersion along in-plane directions for $k_z=0$ ($\Gamma$, M and X points in the Brillouin zone) and $k_z=\pi$ (Z, A, and R point respectively). Out-of-plane dispersion is much smaller than the observed broadening of the Fe $3d$ bands in the region of interest, and therefore does not contribute significantly neither to the width of spectral features nor to the position of experimental peaks. For the case of arsenide $p$ band the part of the band with very little $k_z$ dispersion (band tops at $\Gamma$ and M points) were used to estimate the scattering rate.



%
%
%
%
%